# Flexoelectric effect mediated spin-to-charge conversion at amorphous-Si thin film interfaces

Ravindra G Bhardwaj[1,†], Anand Katailiha[1,†], Paul C. Lou[1], W.P. Beyermann[2] and Sandeep Kumar[1,3,*]

[1] Department of Mechanical Engineering, University of California, Riverside, CA 92521, USA

[2] Department of Physics and Astronomy, University of California, Riverside, CA 92521, USA

[3] Materials Science and Engineering Program, University of California, Riverside, CA 92521, USA

[†] Equal contribution.

Abstract

Interfacial spin to charge conversion arises due to an electric potential perpendicular to the interface. The electric potential can be artificially induced, for example, using ferroelectric and piezoelectric thin films at the interface. An alternate way to induce the electric potential could be flexoelectric field. The flexoelectricity can be observed in all the material that either have or lack inversion symmetry, additionally no large gate bias is needed. In this experimental study, we report large spin to charge conversion (spin-Hall angle- 0.578) at $Ni_{80}Fe_{20}$/amorphous-Si interfaces attributed to flexoelectricity mediated Rashba spin-orbit coupling. The flexoelectricity at the interface also gave rise to interlayer spin-acoustic phonon or flexo-magnetoelastic coupling. In addition to spin-charge conversion, the strained interfaces also led to almost three-fold increase in anomalous Nernst effect. This strain engineering for spin dependent thermoelectric behavior at room temperature opens a new window to the realization of spintronics and spin-caloritronics devices.

## A. Introduction

Structure inversion asymmetry (SIA) at interfaces give rise to Rashba spin-orbit coupling (SOC), which is being extensively investigated for various spintronics devices[1,2]. We hypothesized that the spontaneous polarization under an applied strain gradient (known as the flexoelectric effect – shown in Figure 1 (a)) will give rise to an electric field normal to the interface, which can give rise to Rashba SOC and large spin to charge conversion especially at semiconductor interfaces. For a uniform strain gradient, the flexoelectric field can be written as:

$$E_l = \frac{\mu_{ijkl}}{\varepsilon} \frac{\partial \epsilon_{ij}}{\partial x_k} \tag{1}$$

Where $\frac{\partial \epsilon_{ij}}{\partial x_k}$, $\mu_{ijkl}$ and $\varepsilon$ are strain gradient, flexoelectric coefficient and dielectric constant respectively. Hence, the flexoelectric field will give rise to interfacial Rashba SOC[3] at the interface, which can be written as:

$$H_R \propto (\vec{E} \times \vec{p}) \cdot \vec{\sigma} = E_z(-p_y \sigma_x + p_x \sigma_y) = \frac{\mu_{xx,zz}}{\varepsilon} \frac{\partial \epsilon_{xx}}{\partial x_z}(-p_y \sigma_x + p_x \sigma_y) \tag{2}$$

Where $\vec{p}$ and $\vec{\sigma}$ are momentum and spin polarization vectors respectively. While the flexoelectric polarization is very weak in bulk centosymmetric materials, it can be orders of magnitude larger in nanoscale materials and at interfaces[4–6]. The flexoelectric effect could also lead to charge separation, increasing the charge carrier density in the two-dimension electron gas system (2DES) at the interface. The combined effect of strain[7], strain gradient and charge separation[8] in doped semiconductors will lead to an interfacial 2DES with strong Rashba SOC. Unlike the Rashba SOC in metallic thin film structures, the flexoelectric field driven Rashba SOC can be controlled by changing the strain gradient. Over the years, experimental studies have reported both proximity driven strong spin-splitting at Si surfaces[9,10] as well as weak Rashba SOC in Si 2DES[11,12]. Recently, Yang et al. experimentally reported that flexo-photovoltaic response in p-Si was an order of magnitude larger than $SrTiO_3$ [13]. Similarly, Lou et al. reported observation of large spin-Hall effect in Si due to strain gradient[14]. These studies led us to hypothesize that strain gradient mediated 2DES at the p-doped Si interface might exhibit stronger Rashba SOC and large spin-to-charge conversion without any gate biasing. Our hypothesis is similar to a recent report where ferroelectric field perpendicular to the interface was used to manipulate spin to charge conversion[15]. Instead, we propose to use flexoelectric field, which does not require large gate bias. We chose amorphous-Si (a-Si) as it lacks a center of inversion and may exhibit larger flexoelectric polarization as well as strain splitting at nanoscale. The metal interface with p-doped a-Si will have a 2DES due to band bending as shown in Figure 1 (a). We chose highly doped Si layer to have higher carrier concentration due to charge separation[8]. The hypothesis was tested using a spin-Seebeck effect (SSE) measurement in p-doped a-Si thin film interfaces. The spin-to-charge conversion in the case of a-Si was found to be an order of magnitude larger than that of Pt. This difference disappeared when the strain was relaxed, proving our hypothesis that strain and strain gradient caused by residual stresses give rise to 2DES and Rashba SOC.

### B. Experimental results

The SSE, discovered by Uchida et al.[16,17], is a composite effect of thermal spin current and spin-to-charge conversion, which produces an electric field given by,

$$E_{ISHE} = -S\sigma \times \nabla T \tag{3}$$

where, S is spin Seebeck coefficient and $\sigma$ is the spin polarization vector. In the above equation, $\sigma$ can be replaced by M (magnetization), which gives rise to an equation for an anomalous Nernst effect (ANE). It was noted that both SSE and ANE have identical symmetry behavior, which may lead to false identification of ANE as SSE[18,19]. Hence, we fabricated[20,21](Supplementary Information A) four devices with the following sample structures: a $Ni_{80}Fe_{20}$ (25 nm) control device, a Pt (3 nm)/$Ni_{80}Fe_{20}$ (25 nm), a $Ni_{80}Fe_{20}$ (25 nm)/a-Si (50 nm) and a $Ni_{80}Fe_{20}$ (25 nm)/a-Si (5 nm). These devices will allow us to estimate the ANE contribution and also evaluate the efficiency of the spin-to-charge conversion of the a-Si. The order of layers in the Pt sample was switched to ensure that the anomalous Nernst effect (ANE) and transverse spin dependent thermal responses have the same sign[22]. The schematic of the device and a representative image are shown in Figure 1 (b,c), respectively. The temperature gradient ($\Delta T_z$) was generated across the thickness of the thin film sample by passing an electric current (I) through the Pt heater, as shown in Figure 1 (b). The residual compressive stresses, due to lattice mismatch and thermal expansion[23], in the MgO and Pt heater layer (Supplementary Figure S1) lead to a strain as well as a strain gradient in the sample, as shown in the Figure 1 (d). The resulting flexoelectric effect will lead to the hypothesized Rashba SOC at the interface. It was noted that the Pt heater layer was electrically isolated from the sample and will make no contribution towards any spin dependent response. The structure of the thin films was verified using a high-resolution transmission electron microscope (HRTEM) study (Supplementary information-A) for the $Ni_{80}Fe_{20}$ (25 nm)/a-Si (5 nm) sample, as shown in Figure 1 (e)-(f). From the HRTEM study, we observed a continuous a-Si thin film layer for a 5 nm a-Si sample. We did not observe any measurable Ni or Fe diffusion in the a-Si layer as shown in Figure 1 (f). Using HRTEM and AFM measurement, we deduced that the mean roughness for the a-Si samples is ~1.22 nm, as shown in Supplementary Figure S2.

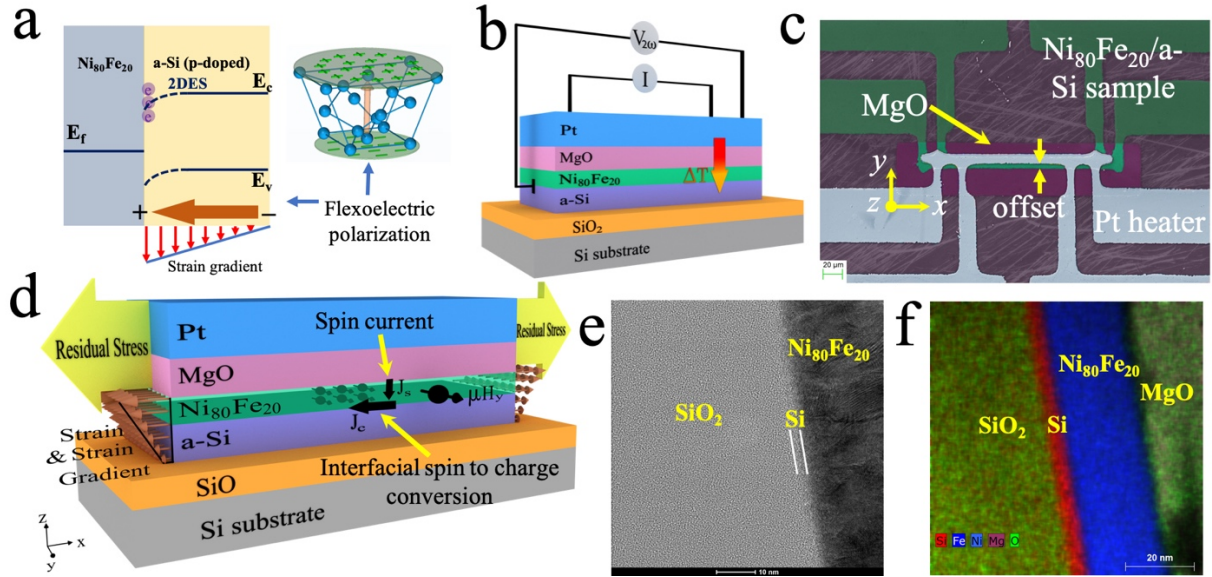

Figure 1. (a) a schematic showing flexoelectric polarization in Si lattice due to strain gradient and band diagram showing the mechanistic origin of Rashba SOC and interlayer coupling, (b) a schematic showing the experimental setup with the temperature gradient, (c) a representative false color scanning electron micrograph showing the experimental device, (d) a schematic showing the origin of strain and strain gradient leading to the interfacial spin-Seebeck effect in our experimental setup (e) a high resolution transmission electron micrograph showing the layered structure of the experimental specimen and (f) an energy dispersive X-ray spectroscopy elemental map showing the thin film layers and interfaces.

The transverse thermoelectric response measurements were undertaken at 30 mA/5 Hz of heating current and resulting $V_{2\omega}$ response (being quadratic in heating current) was acquired as a function of the magnetic field (1500 Oe to -1500 Oe) applied in the y-direction (normal to the temperature gradient). This is called in-plane magnetized (IM) configuration of thermoelectric response measurement. The $V_{2\omega}$ responses were measured at 300 K as shown in Figure 2 (a) and the measured responses are listed in Table 1. All the samples had the resistances between 320 Ω-350 Ω and $Ni_{80}Fe_{20}$ thin film resistivity was expected to be $5\times10^{-7}$ Ωm. We used the 3ω method[24] to estimate the increase in heater temperature and finite element method simulation to estimate the temperature drop across $Ni_{80}Fe_{20}$ layers as shown in Supplementary section C and Supplementary Figure S3-S4. The heater temperature was stable and did not deviate under an applied magnetic field as shown in Supplementary Figure S5. In the case of $Ni_{80}Fe_{20}$, the measured response of 15.1 µV was attributed to the ANE (easy axis). In the case of the Pt sample, the estimated transverse spin dependent response was due to SSE having magnitude of ~18 µV; a behavior consistent with the reported ANE and SSE responses for the Pt/ $Ni_{80}Fe_{20}$ sample[22]. The corresponding transverse spin dependent thermal response in the case of the a-Si samples will be ~70 µV and 150.35 µV for 50 nm and 5 nm a-Si, respectively, which was 4 times and 8.5 times that of SSE response in Pt sample. This difference cannot arise due to shunting effect since the resistivity of the a-Si is two order of magnitude larger than $Ni_{80}Fe_{20}$ thin film. The interfacial roughness between 50 nm and 5 nm a-Si samples was not different, as shown in Supplementary Figure S2, and will not cause observed difference in measurements. The estimated the spin-Seebeck coefficients were 1.032±0.1 µV/K and 0.458±0.05 µV/K for 5 nm a-Si and 50 nm a-Si devices, respectively, (details in Supplementary Information C). The corresponding ANE (easy axis) coefficient of $Ni_{80}Fe_{20}$ was estimated to be 0.1 µV/K. This value was approximately two times larger than the values reported in literature (0.045 µV/K)[25], which could be due to interfacial (oxide) contribution.

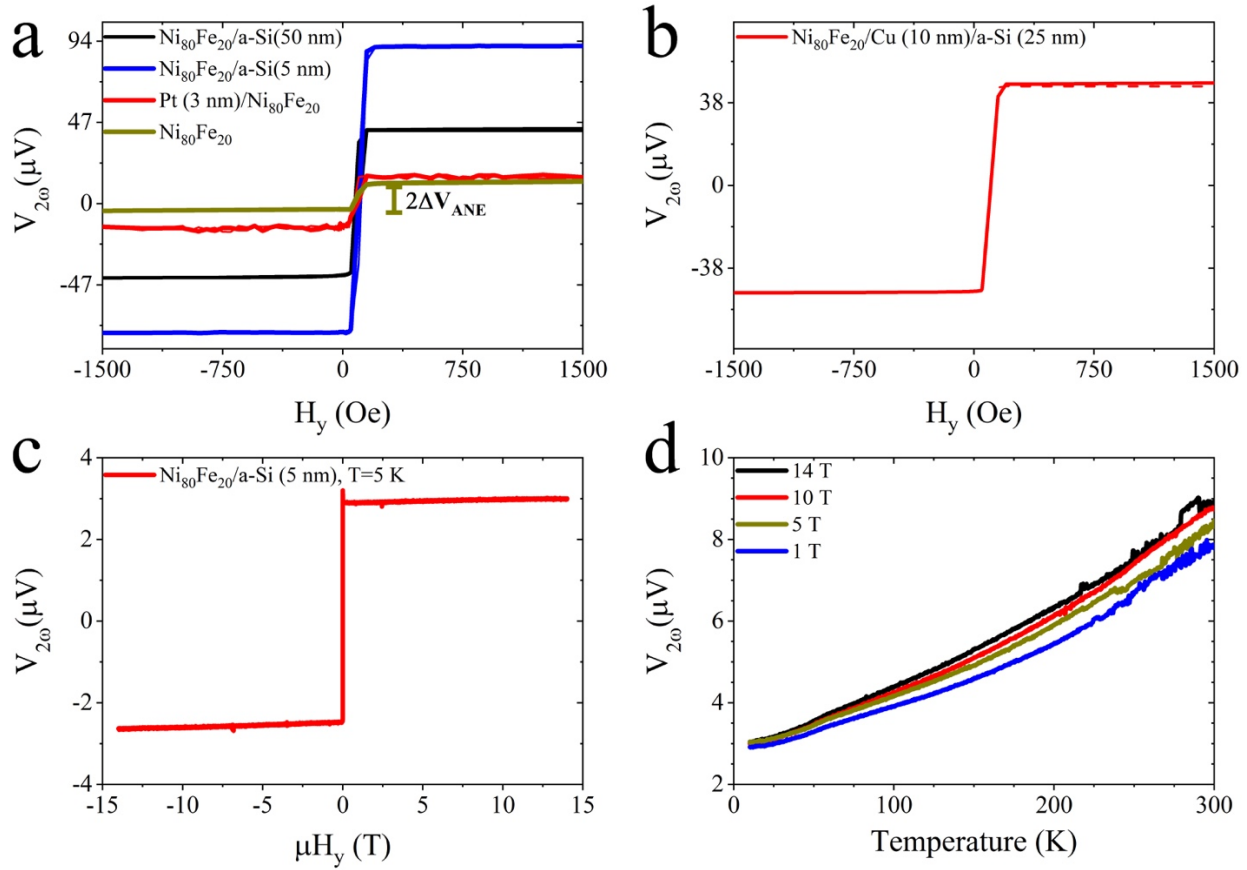

Figure 2. (a) the transverse spin dependent thermal measurement for $Ni_{80}Fe_{20}$ (25 nm), Pt (3 nm)/$Ni_{80}Fe_{20}$ (25 nm), $Ni_{80}Fe_{20}$ (25 nm)/a-Si (50 nm) and $Ni_{80}Fe_{20}$ (25 nm)/a-Si (5 nm) samples at 300 K, (b) the spin-Seebeck effect measurements for $Ni_{80}Fe_{20}$ (25 nm)/Cu (10 nm)/a-Si (25 nm) sample at 300 K, (c) the high magnetic field transverse spin dependent thermal measurement for $Ni_{80}Fe_{20}$ (25 nm)/a-Si (5 nm) sample at 5 K, and (d) the transverse spin dependent thermal measurement as a function of temperature for $Ni_{80}Fe_{20}$ (25 nm)/a-Si (5 nm) sample at an applied magnetic field ($\mu H_y$) of 1 T, 5 T, 10 T and 14 T from 300 K to 10 K. The fluctuations in the temperature-dependent measurements are due to instrumental settings.

We also measured the transverse thermoelectric response of 33.5 μV in 50 nm a-Si sample at 20 mA of heater current as shown in Supplementary Figure S6 (a). Heating power follows a quadratic relationship with the current. And corresponding response at 30 mA should have been 75.37 μV (2.25 times 33.5 μV) as compared to 85 μV. This difference was attributed to the additional strain gradient due to thermal expansion mismatch at the interface. Similarly, we measured the angle dependent $V_{2\omega}$ response at 20 mA of heating current in sample with 5 nm a-Si as shown in Supplementary Figure 6 (b). The measurement showed cosine behavior as expected for ANE/SSE behavior. The overall response was measured to be ~67 µV. Similar to 50 nm sample, the response at 30 mA was larger than expected, which was again attributed to the strain gradient. The larger strain gradient was also responsible for larger transverse spin dependent thermal response in the thinner 5 nm a-Si sample. Then, we introduced a 10 nm Cu layer in between the $Ni_{80}Fe_{20}$ (25 nm) and a-Si (25 nm) to remove the ferromagnetic proximity effect[26,27] in the strained configuration. For similar heating power, we measure the $V_{2\omega}$ response to be 95.8 µV for this, which was larger than that of Pt sample. The addition of Cu layer should have reduced the transverse thermal response due to shunting effect. However, the larger response in this case might arise from an additional strain gradient in 25 nm a-Si layer due to additional layer. This measurement eliminated the ferromagnetic proximity effect.

Table 1. List of ANE (easy axis) and SSE responses and corresponding coefficients for both unstrained and strained samples in IM configuration.

| IM configuration | Sample | $2(V_{ANE} + V_{SSE})$ (µV) | ANE/SSE Coefficient (µV/K) |
|---|---|---|---|
| Strained | $Ni_{80}Fe_{20}$ | 15.1 | 0.1 (ANE) |
| | Pt/$Ni_{80}Fe_{20}$ | 33 | 0.117 (SSE) |
| | $Ni_{80}Fe_{20}$/a-Si (50 nm) | 85 | 0.458±0.05 (SSE) |
| | $Ni_{80}Fe_{20}$/a-Si (5 nm) | 165.45 | 1.032±0.1 (SSE) |
| | $Ni_{80}Fe_{20}$/Cu/a-Si (25 nm) | 95.8 | - |
| Unstrained | $Ni_{80}Fe_{20}$ | 0.35 | - |
| | Pt/$Ni_{80}Fe_{20}$ | 0.71 | - |
| | a-Si (50 nm)/$Ni_{80}Fe_{20}$ | 0.95 | - |

To uncover the effects of strain gradient, we fabricated a set of control devices with unstrained samples where the position of sample and heater were switched as shown in Supplementary Figure S7. The transverse thermoelectric responses in IM configuration were measured in a-Si (50 nm)/$Ni_{80}Fe_{20}$, $Ni_{80}Fe_{20}$ and Pt/$Ni_{80}Fe_{20}$ unstrained samples, as shown in Supplementary Figure S8 (a)-(d) and listed in Table 1. The spin dependent thermal response in a-Si sample was found to be 1.67 times larger than that of Pt sample as compared to four times in strained sample (Details in Supplementary Section E). This difference clearly supported our hypothesis of strain gradient mediated Rashba SOC at the interface. The temperature information in unstrained sample was not measured and coefficient could not be calculated.

Then, we measured the $V_{2\omega}$ response at 5 K for 10 mA of heating current and applied magnetic field from 14 T to -14 T in the 5 nm a-Si bilayer sample, as shown in Fig. 2 (c). The magnonic spin current could be suppressed at low temperatures and with high magnetic field[28,29], which we did not observe as shown in Figure 2 (c). This suggested that the origin of the spin current was due to the spin dependent Seebeck effect (SDSE). This assertion was further supported by the $V_{2\omega}$ response as a function of temperature from 10 K to 300 K with applied magnetic field 1 T, 5 T, 10 T, and 14 T as shown in Fig. 2 (d). The $V_{2\omega}$ response increased as the magnetic field was increased. This shows that the origin of spin current was most likely electronic and not magnons[30], hence the SDSE was the underlying cause of the transverse spin dependent thermal response reported in this work. The increase in the $V_{2\omega}$ response with increasing magnetic field could be attributed to an increase in SDSE due to a reduction in electron-magnon scattering at higher magnetic fields. While the origin of spin current was due to SDSE, we propose to call transverse spin dependent thermal response as SSE since the detection was attributed to the spin-to-charge conversion.

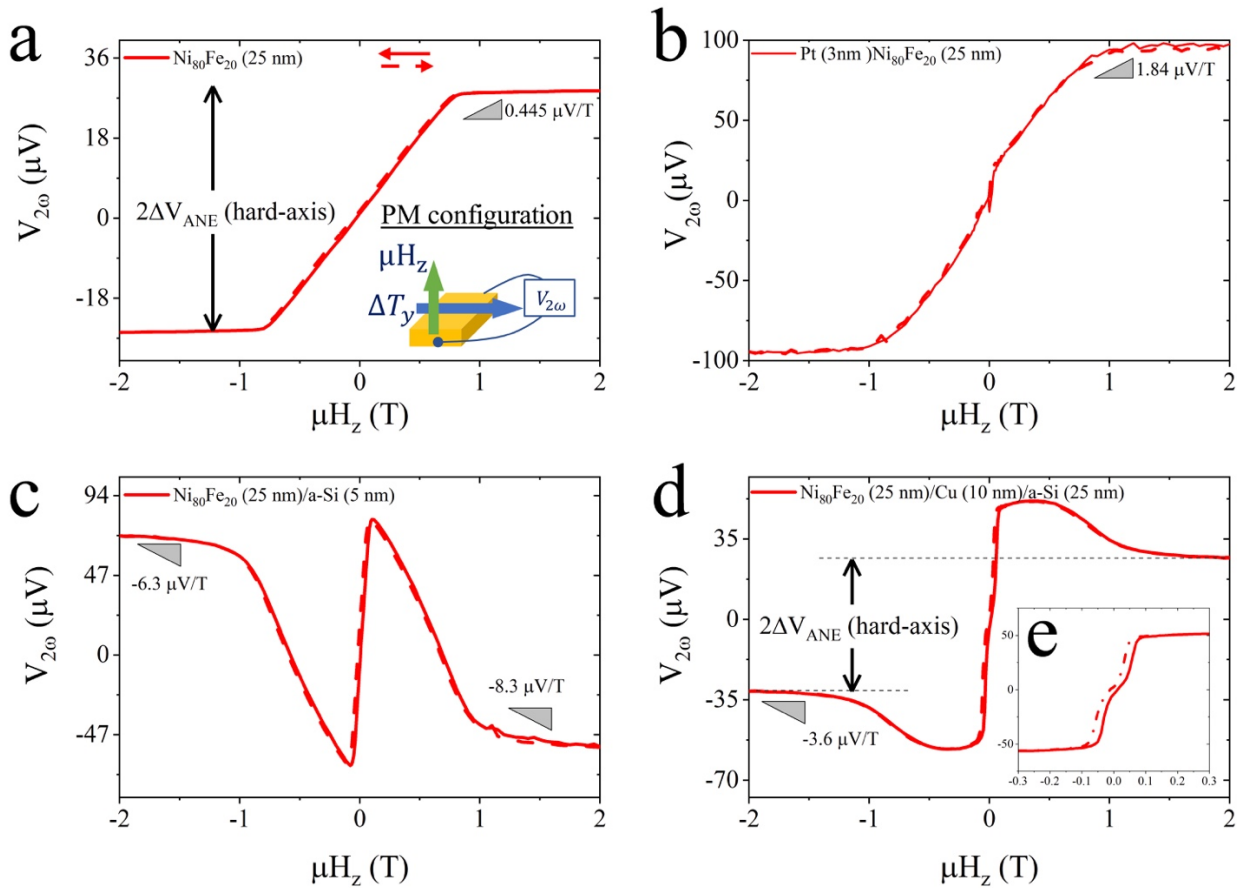

Figure 3. The $V_{2\omega}$ response perpendicularly magnetized (PM) configuration for an applied magnetic field sweep from 2 T to -2 T in (a) $Ni_{80}Fe_{20}$ (25 nm), (b) Pt(3nm)/$Ni_{80}Fe_{20}$ (25 nm).

Similar measurement in (c) $Ni_{80}Fe_{20}$ (25 nm)/a-Si (5 nm) and (d) $Ni_{80}Fe_{20}$ (25 nm)/Cu (10 nm)/a-Si (25 nm) samples. (e) shows the low field behavior (between 0.3 T to -0.3 T) in $Ni_{80}Fe_{20}$ (25 nm)/Cu (10 nm)/a-Si (25 nm) sample. Inset in (a) shows the measurement setup. Arrows in (a) shows the direction of magnetic field sweep.

Then, we undertook measurements in perpendicularly magnetized (PM) configuration[19]. In PM configuration, in-plane temperature difference ($\Delta T_y$) and out of plane magnetic field ($\mu H_z$) led to transverse thermal response as shown in inset of Figure 3 (a). In the measurement devices, the heater and sample had an offset due to lithographic misalignment, as shown in Figure 1 (c). This offset gave rise to a temperature gradient along the y-axis. Since the misalignment was not controlled, the resulting temperature difference varies from sample to sample. The transverse thermoelectric responses in all the unstrained and strained were measured in PM configuration for applied magnetic field of 2 T to -2 T as shown in Supplementary Figure S9 and Figure 3, respectively, and are listed in Table 2. We also estimated the ANE (hard axis) coefficients, ONE responses and ONE coefficients for all the samples as shown in Table 2. The temperatures were estimated using planar Nernst effect (PNE) coefficient of 70 nV/K[31] and assuming ONE coefficient will not change in $Ni_{80}Fe_{20}$ sample since it is bulk property (details in Supplementary Section E).

Table 2. List of ANE (hard axis) and ONE responses and corresponding coefficients for both unstrained and strained samples in PM configuration.

| PM configuration | Sample | $2V_{ANE}$ (Hard-axis) (µV) | ANE Coefficient (µV/K) | ONE (µV/T) | ONE coefficient (µV/(KT)) |
|---|---|---|---|---|---|
| **Strained** | $Ni_{80}Fe_{20}$ | 52.50 | 6.6 | 0.445 | 0.113 |
| | $Pt/Ni_{80}Fe_{20}$ | 185 | 4.65 | 1.84 | 0.0926 |
| | $Ni_{80}Fe_{20}$/a-Si (5 nm) | 95.93 | 6.6 | -6.3 (-µH$_z$) and -8.3 (+µH$_z$) | -0.542 (-µH$_z$) and -0.714 (+µH$_z$) |
| | $Ni_{80}Fe_{20}$/Cu/a-Si (25 nm) | 71.5 | 6.6 | -3.6 µV/T | -0.664 |
| **Unstrained** | $Ni_{80}Fe_{20}$ | 15.45 | 2.565 | 0.278 | 0.113 |
| | $Pt/Ni_{80}Fe_{20}$ | 15.40 | 2.565 | 0.338 | 0.0926 |
| | a-Si (50 nm)/$Ni_{80}Fe_{20}$ | 2.5 | 2.565 | 0.0015 | 0.0015 |

The transverse thermoelectric response behavior in strained configuration are shown in Figures 3 (a)-(d). The ANE (hard-axis) coefficients were estimated to be 2.565 (±10%) µV/K and 6.6(±10%) µV/K for unstrained and strained $Ni_{80}Fe_{20}$ samples, respectively. The ANE (hard-axis) coefficient was found to be ~0.128 times the Seebeck coefficient in the unstrained sample and it was similar to bulk value of 0.13[32,33], which then increased to 0.33 in the strained sample (Supplementary Section E). This increase arose due to self-induced spin dependent behavior at strained MgO and $SiO_2$ interfaces[34]. This argument was supported by reduction in ANE (hard-axis) coefficient (4.65 µV/K) in strained $Pt/Ni_{80}Fe_{20}$ sample, where $Ni_{80}Fe_{20}$ layer had only one interface with the oxide (details in Supplementary Section E). This measurement showed that strained interfaces could change the behavior significantly. This measurement also showed that there was an order of magnitude difference between easy-axis and hard-axis ANE coefficients. This was attributed to the increased surface/interfacial scattering in the out of plane temperature gradient causing reduction in easy-axis ANE coefficient. The $S_{ONE}$ were 0.113(±10%) µV/(KT) and 0.0926(±10%) µV/(KT) in $Ni_{80}Fe_{20}$ and $Pt/Ni_{80}Fe_{20}$ samples, respectively, as shown in Table 2.

The transverse thermoelectric response in $Ni_{80}Fe_{20}$/a-Si (5 nm) and $Ni_{80}Fe_{20}$/Cu/a-Si (25 nm) samples had ANE (hard axis) response, ONE response and an additional unknown response superimposed on them as shown in Figure 3 (c,d). The ONE response was negative for both the samples. The ONE response from impurity scattering is positive and from acoustic phonon scattering is negative. Hence, this transition was attributed to the acoustic phonon scattering[35]. However, the change in sign of ANE (hard axis) response in the $Ni_{80}Fe_{20}$/a-Si (5 nm) bilayer was intriguing since ANE in $Ni_{80}Fe_{20}$ was considered to be scattering independent[36]. Yang et al.[37] observed a similar to sign in anomalous Hall effect (AHE) in $Mn_xSi_{1-x}$, which they attributed to Rashba SOC. However, Rashba SOC might not lead to change in sign directly. The AHE due to skew-scattering is considered to have opposite sign as compared to intrinsic AHE[38-40]. We propose that the change in sign of ANE (hard axis) was due to transition from intrinsic mechanism to skew-scattering mechanism. We propose that acoustic phonon mediated flexo-magnetoelastic interlayer coupling between $Ni_{80}Fe_{20}$ and a-Si was the underlying cause of skew-scattering and it was also supported by negative ONE response from acoustic phonon scattering. The response in unstrained a-Si (50 nm)/$Ni_{80}Fe_{20}$ corroborated our hypothesis since both ONE and ANE (hard axis) responses were positive as shown in Table 2 and Supplementary Figure S9 (c). The positive ANE (hard axis) response in sample with Cu interlayer also supported our hypothesis as shown in Table 2 and Figure 3 (d).

These measurements further supported our contention that the response in IM configuration was due to strain gradient mediated SSE and not ANE (easy axis). The ANE (hard axis) response in the PM configuration was negative whereas the corresponding SSE response (in IM configuration) was positive in $Ni_{80}Fe_{20}$/a-Si (5 nm) sample. If the origin of the thermoelectric response in $Ni_{80}Fe_{20}$/a-Si (5 nm) sample in IM configuration was ANE (easy axis) then the sign should have been negative as opposed to positive. These measurements in PM configuration also suggested an additional reason for the larger SSE response in sample with Cu interlayer. In a-Si samples without the Cu interlayer, the ANE (easy axis) response was, possibly, negative and should be added to the measured response whereas in sample with Cu spacer the ANE (easy axis) response was positive and already included in the total response. If we use this analogy then the SSE response in 50 nm a-Si, 5 nm a-Si and sample with Cu spacer should be 100.1 µV, 180.45 µV and 80.7 µV, respectively. This might explain the absence of shunting effect in the raw data in addition to extra strain from Cu layer as stated earlier.

Then, we analyzed the unknown low field response in a-Si samples with and without Cu interlayer. When an external magnetic field perpendicular to the Rashba interface is applied, it will induce an out of plane spin component in Rashba 2DES[41]. This was proposed to be the cause of observed low field response since large magnetic field quenched the observed response. In $Ni_{80}Fe_{20}$/a-Si (5 nm) sample, the ONE response was significantly different for positive (-8.3 µV/T) and negative (-6.3 µV/T) magnetic fields. This asymmetry might arise due to interfacial spin polarization (out of plane) and was not observed in any other sample. In addition, the existence of out of plane spin component was verified from the low field response (in PM configuration) in sample with Cu interlayer, which exhibited a switching behavior that was similar to an exchange biased (between Si/Cu and $Ni_{80}Fe_{20}$/Cu interfaces) layered thin film as shown in Figure 3 (e). This exchange biased switching was absent in the IM configuration as shown Figure 2(b) – a behavior expected of Rashba coupled 2DES. This low field behavior was clearly absent in $Ni_{80}Fe_{20}$ and Pt/$Ni_{80}Fe_{20}$ samples as shown in Figure 3(a,b), respectively. In addition, this low field behavior was not observed in unstrained a-Si (50 nm)/$Ni_{80}Fe_{20}$ sample as shown in Supplementary Figure S9 (c), which meant that strain gradient was expected to be the primary cause of observed responses in strained a-Si samples shown in Figure 3 (c,d).

C. **Discussion**

The experimental results presented in this study demonstrated large spin-Seebeck responses. The next step was quantitative description of the spin-to-charge conversion efficiency. The spin-Hall angle of the Pt was 0.068[42] for $Ni_{80}Fe_{20}$ ferromagnetic source. We estimated that the SSE response of 5 nm a-Si was 8.5 times that of SSE in Pt sample. The corresponding spin-Hall angle for 5 nm a-Si will be 0.578. This value was much larger than that of most heavy metals[43]. The effective spin mixing conductance in bulk Si ($1.74 \times 10^{19}$ $m^{-2}$ – $5.2 \times 10^{19}$ $m^{-2}$)[44,45] was reported to be similar to that of Pt ($2.1 \times 10^{19}$ $m^{-2}$) [46]. Hence, the spin injection efficiency might not be the underlying cause of large spin to charge conversion in a-Si. However, the strained a-Si might have larger spin mixing conductance due to magnetoelastic coupling and might contribute towards larger SSE response. The largest spin-Seebeck coefficient in this work was estimated to be 1.032 µV/K, which was smaller than YIG/NiO/Pt system reported by Lin et al[47]. Our value was similar to $Fe_3O_4$/Pt system[48]. However, we achieved it using strain gradient and in spite of insignificant SOC in Si.

For Rashba SOC-coupled 2DES, an alternate method to define the spin-to-charge conversion efficiency is Rashba-Edelstein length $\lambda_{IEE} = \alpha_R \tau / \hbar$, where $\alpha_R$ and $\tau$ are Rashba parameter and relaxation time, respectively. The Rashba spin splitting cannot be calculated from transport measurements. However, the Rashba-Edelstein length can be estimated using multiplication of spin-Hall angle and spin diffusion length. The smallest thickness of the a-Si used in our studies was 5 nm and it had the highest spin-Seebeck response. We assumed that spin diffusion length is approximately 5 nm. Using this value, we estimated the Rashba-Edelstein length for a-Si samples to be ~2.68 nm. This value was larger than the value of Rashba-Edelstein length in the case of strained HgTe [49]($\lambda_{IEE} = 2$) and α-Sn[50] ). The largest magnitude of Rashba-Edelstein length has been reported for the 2DES at LAO/STO interface under a large gate bias (125 V[51] to 175 V[52]). However, there was no need for gate bias in the case of p-doped a-Si, as shown in this study. The Rashba parameter was estimated to be $1.3 \times 10^{-12}$ eV-m when compared with LAO/STO 2DES system ($\lambda_{IEE} = 6.4$ nm) [51].

D. **Conclusion**

In conclusion, we reported one of the largest thermal spin-to-charge conversion response at a strained metal/a-Si interface. The efficiency of spin-to-charge conversion was an order of magnitude larger than that of Pt and similar to topological insulator surface states, however, without any added fabrication complexity. We also presented the first experimental proof of flexoelectric polarization mediated Rashba SOC, which avoided any gate biasing. The flexoelectric field also induced spin-acoustic phonon (magnetoelastic) coupling that inverts the sign of ordinary Nernst effect as well as anomalous Nernst effect from positive to negative in $Ni_{80}Fe_{20}$/a-Si bilayers. Interfacial strain not only gave rise to large spin to charge conversion but it also led to 2.5 times increase in anomalous Nernst coefficient in $Ni_{80}Fe_{20}$ thin films. The observation of a large SSE in an a-Si challenges the inherent need for large intrinsic SOC in spintronics and spin-caloritronics research.

**Acknowledgement**

The fabrication of experimental devices was done at the Center for Nanoscale Science and Engineering at UC Riverside. The electron microscopy sample preparation and imaging were done at the Central Facility for Advanced Microscopy and Microanalysis at UC Riverside. SK acknowledges research gift from Dr. S Kumar.

**Conflict of interest**

Authors declare no conflict of interest.

**Supplementary information- Flexoelectric effect mediated spin to charge conversion at amorphous-Si thin film interfaces**

Ravindra G Bhardwaj[1], Anand Katailiha[1], Paul C. Lou[1], W.P. Beyermann[2] and Sandeep Kumar[1,3,*]

[1] Department of Mechanical Engineering, University of California, Riverside, CA 92521, USA

[2] Department of Physics and Astronomy, University of California, Riverside, CA 92521, USA

[3] Materials Science and Engineering Program, University of California, Riverside, CA 92521, USA

### A. Device fabrication process and experimental measurement

We take a prime Si wafer and deposit 350 nm of thermal silicon oxide using chemical vapor deposition (CVD). Using lift-off photolithography, we then deposit the sample to be studied using the RF sputtering. The sputtering deposition will have substrate conformal thin film coating. Hence, it will have the same interfacial and surface roughness as the underlying layers. The p-Si target used to deposit amorphous-Si layer is Boron-doped with resistivity of 0.005-0.01 $\Omega$-cm. The second lift-off photolithography is carried out to deposit 50 nm MgO to electrically isolate the sample from the heater. The third lift-off photolithography is then used to deposit heater composed of Ti (10 nm)/Pt (100 nm) using e-beam evaporation.

The residual stresses from Pt (heater) and MgO (insulator) layers are proposed to be the underlying cause of strain gradient mediated Rashba SOC. The existence of residual stresses can be seen in Supplementary Figure S1, where the Pt heater was delaminated due to the residual stresses.

For the second set of devices (unstrained) with switched sample and heater positions, we first deposited Ti (10 nm) / Pt (100 nm) on a Si wafer with predisposition of thermal silicon oxide (650 nm) using CVD. We then sputter 50 nm MgO using RF sputtering for electrical isolation. We fabricated two set of devices having the a-Si (50 nm)/$Ni_{80}Fe_{20}$ (25 nm) bilayer sample and $Ni_{80}Fe_{20}$ (25 nm) sample on top of the MgO.

All the measurements were carried out at 30 mA of heating current except for strained $Ni_{80}Fe_{20}$/Pt device where measurement was carried out at 25 mA. However, the measured response at 25 mA was multiplied with a factor of 1.44 to get the equivalent response to 30 mA. This was done to protect the device and to have a complete measurement. The $Ni_{80}Fe_{20}$/Pt devices were found to be fragile for larger current, which could be due to larger residual stresses. The $V_{2\omega}$ response was measured using lock-in technique.

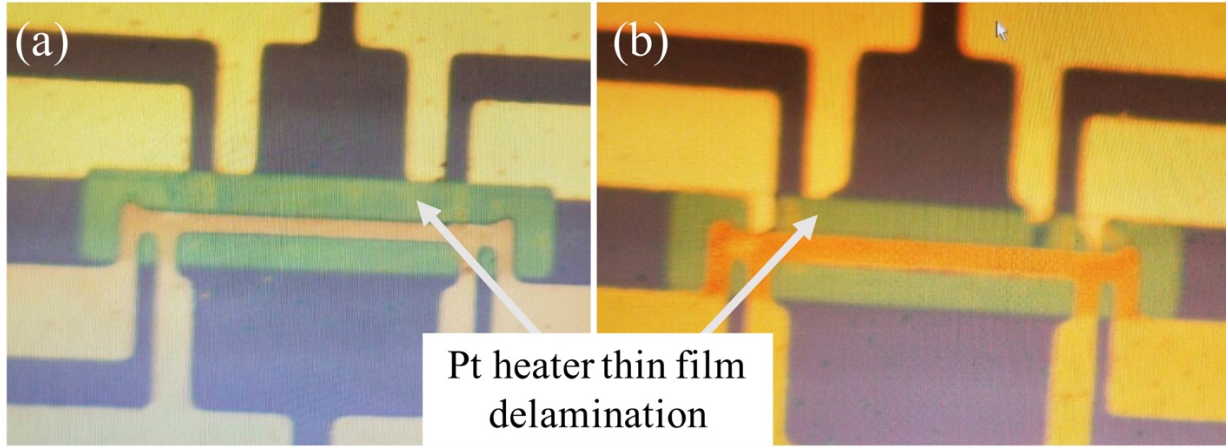

Supplementary Figure S1. (a)-(b)The optical images showing Pt thin film heater layer peeled off (delaminated) due to residual stresses for two devices. The residual stress in Pt heater and MgO layer is proposed to be the cause of strain and strain gradient in the underlying sample.

### B. Sample characterization

<u>TEM sample preparation</u>- TEM lamellae were prepared from the layered sample following established procedures with a Dual Beam scanning electron microscope and FIB instrument using Ga ion source (Quanta 200i 3D, Thermo-Fisher Scientific). First, a strap of 5 μm thick protective Carbon layer was deposited over a region of interest using the ion beam. Subsequently approximately 80 nm thin lamella of was cut and polished at 30 kV and attached to a TEM grid using in-situ Omniprobe manipulator. To reduce surface amorphization and Gallium implantation final milling at 5 kV and 0.5 nA was used to thin the sample further.

<u>S/TEM imaging and analysis</u>- TEM and STEM imaging was performed at 300 kV accelerating voltage in a Thermo-Fisher Scientific Titan Themis 300 instrument, fitted with X-FEG electron source, 3 lens condenser system and S-Twin objective lens. High-resolution TEM images were recorded at resolution of 2048x2048 pixels with a FEI CETA-16M CMOS digital camera with beam convergence semi-angle of about 0.08 mrad. STEM images were recorded with Fischione Instruments Inc. Model 3000 High Angle Annular Dark Field (HAADF) Detector with probe current of 150 pA, frame size of 2048x2048, dwell time of 15 μsec/pixel, and camera length of 245 mm. Energy dispersive X-ray Spectroscopy (EDS) analyzes and elemental mapping were obtained in the STEM at 300 kV, utilizing Thermo-Fisher Scientific SuperX system equipped with 4x30mm$^2$ window-less SDD detectors symmetrically surrounding the specimen with a total collection angle of 0.68 srad, by scanning the thin foil specimens. Elemental mapping was performed with an electron beam probe current of 550 pA at 1024 x1024 frame resolution.

Atomic force microscope (AFM) characterization of surface roughness- The surface roughness of the bilayer sample directly reflects the underlying interfacial roughness. The interfacial roughness cannot be more than the surface roughness since the sputter coating is conformal. The AFM measurements are carried out on samples having 50 nm a-Si and 5 nm a-Si layers as shown in Supplementary Figure S2.

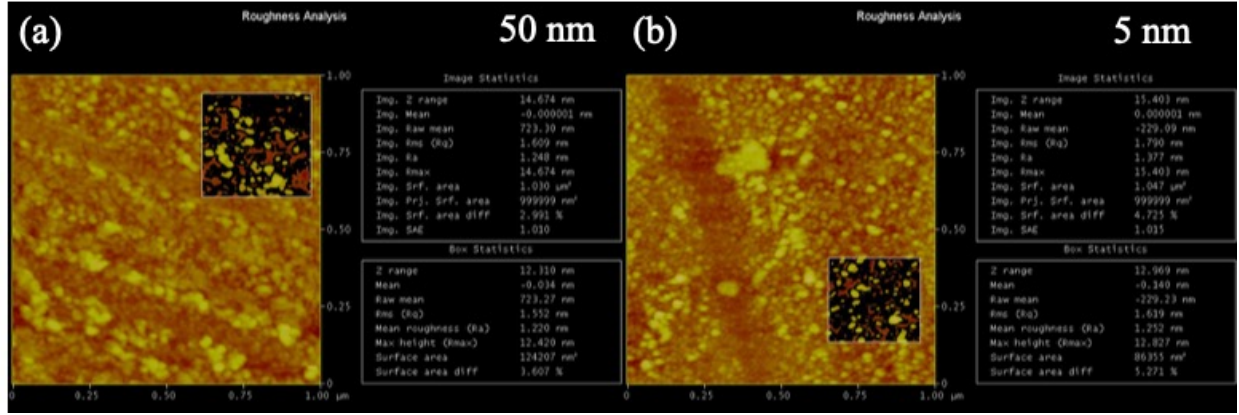

Supplementary Figure S2. The AFM measurements at the surface of Ni$_{80}$Fe$_{20}$ layer in (a) 50 nm and (b) 5 nm a-Si bilayer samples. The mean roughness of both samples is ~1.2 nm.

### C. The temperature estimation and spin-Seebeck coefficient calculation

The increase in temperature across a current carrying heater (Pt) can be estimated using 3ω method, which is given by following equation:

$$\Delta T = \frac{4V_{3\omega}}{R' I_{rms}} \tag{S1}$$

where, $V_{3\omega}$ is the third harmonic response, $R'$ is the resistance as a function of temperature, and $I_{rms}$ is the heating current. The 3ω measurement was carried out at 20 mA and resulting temperature rise was estimated to be 18.817 K and 18.18 K for 50 nm a-Si and 5 nm a-Si devices, respectively. The $V_{3\omega}$ responses were 6.58 mV and 5.0718 mV for 50 nm a-Si and 5 nm a-Si devices, respectively. Similarly, the values of $R'$ were estimated to be 0.07 Ω/K and 0.055 Ω/K for Pt thin film in 50 nm a-Si and 5 nm a-Si devices, respectively. Resistivity of the Pt heater is approximately $3.8 - 4\times10^{-7}$ Ωm. The temperature rise could also be verified using change in resistance due to large current. From a different 5 nm a-Si sample, we measured a change in resistance due to 20 mA of current to be ~1.2 Ω. Consequently, the temperature rise at the heater will be 21.8 K, which was slightly higher and this difference could arise due to small variation in the value of $R'$. Hence, the heater temperature estimates were using $3\omega$ method were correct. The corresponding temperature rise for 30 mA of heating current would be 42.34 K and 41.50 K, respectively. Using this temperature information, we did finite element simulation using COMSOL and found the temperature difference across the Py film for 50 nm a-Si and 5 nm a-Si devices to be 11.924 mK and 11.388 mK, respectively. The values of thermal conductivities used in COMSOL simulations are 20 W/mK, 30 W/mK and 15 W/mK for $Ni_{80}Fe_{20}$, 50 nm a-Si and 5 nm a-Si, respectively. The vertical temperature distribution for 5 nm a-Si and 50 nm a-Si along with temperature data at the interfaces is shown in the Supplementary Figure S3 and S4, respectively.

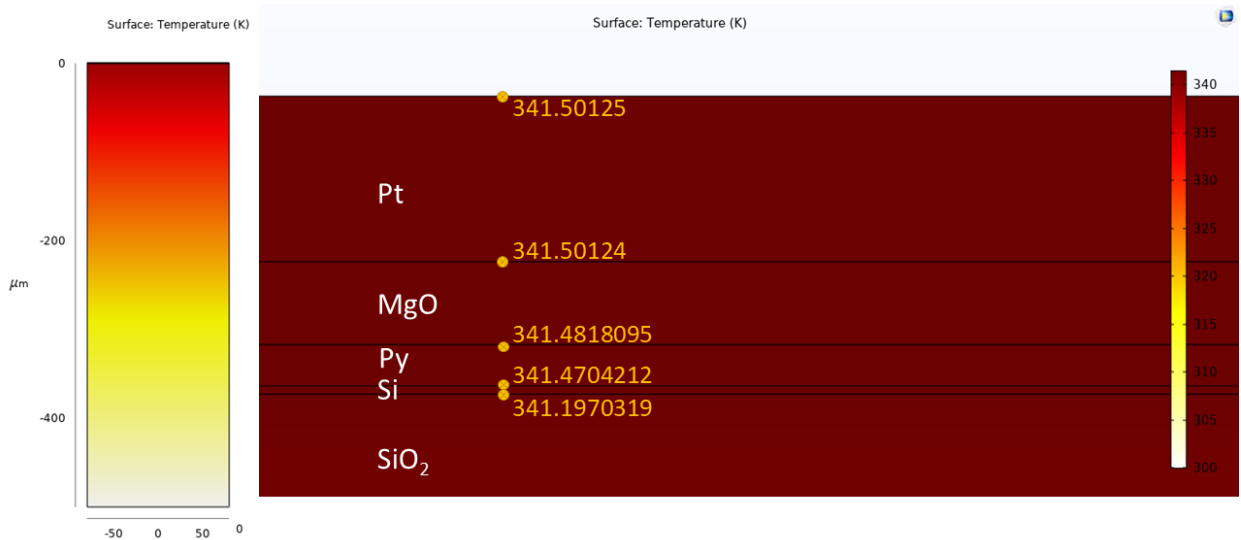

Supplementary Figure S3. Temperature distribution for 5 nm a-Si SSE device

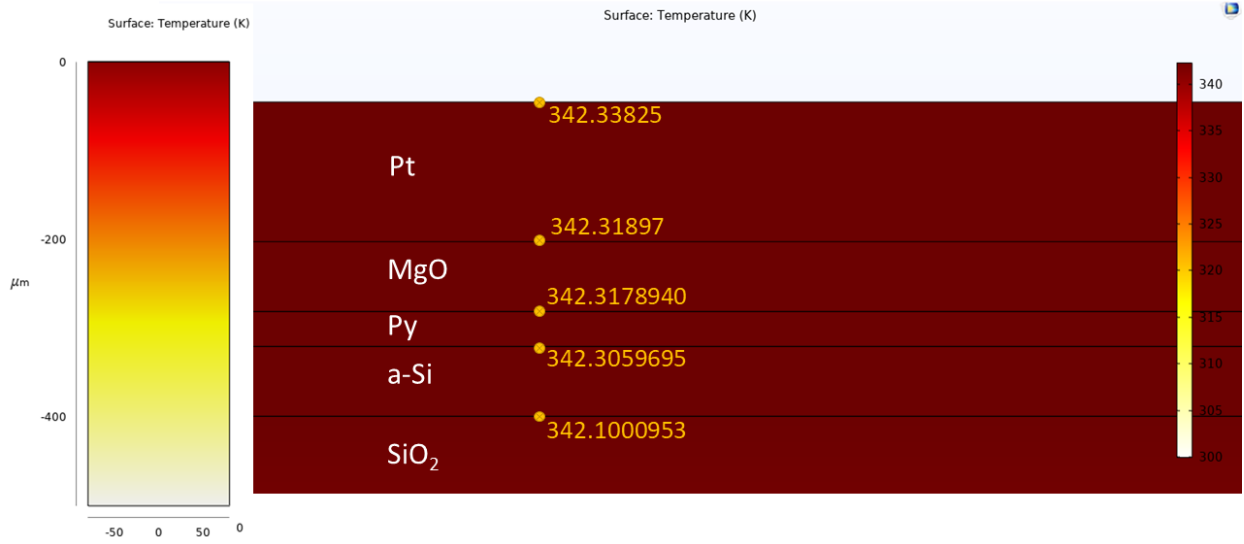

Supplementary Figure S4. Vertical temperature distribution for 50 nm a-Si SSE device

The spin Seebeck coefficient are estimated using following equation:

$$S_{LSSE} = \frac{V_{SSE} t}{L \Delta T} \quad (S2)$$

where, $2V_{SSE} = V_{2\omega}$, t = 25 nm, L = 160 μm and ΔT is the temperature difference across $Ni_{80}Fe_{20}$ film. For 5nm a-Si and 50 nm a-Si the $V_{2\omega}$ is 150.45 μV and 70 μV respectively. So, the $S_{LSSE}$ for 5 nm a-Si and 50nm a-Si is 1.032±0.1 μV/K and 0.458±0.05 μV/K, respectively.

To demonstrate the effect of magnetic field on heater temperature, we measured the angle dependent $V_{3\omega}$ response for an applied magnetic field of 2 T in yx-plane for $Ni_{80}Fe_{20}$/a-Si (5 nm) sample. We did not observe significant drift in the heater temperature at 20 mA of heater current as shown in Supplementary Figure S5.

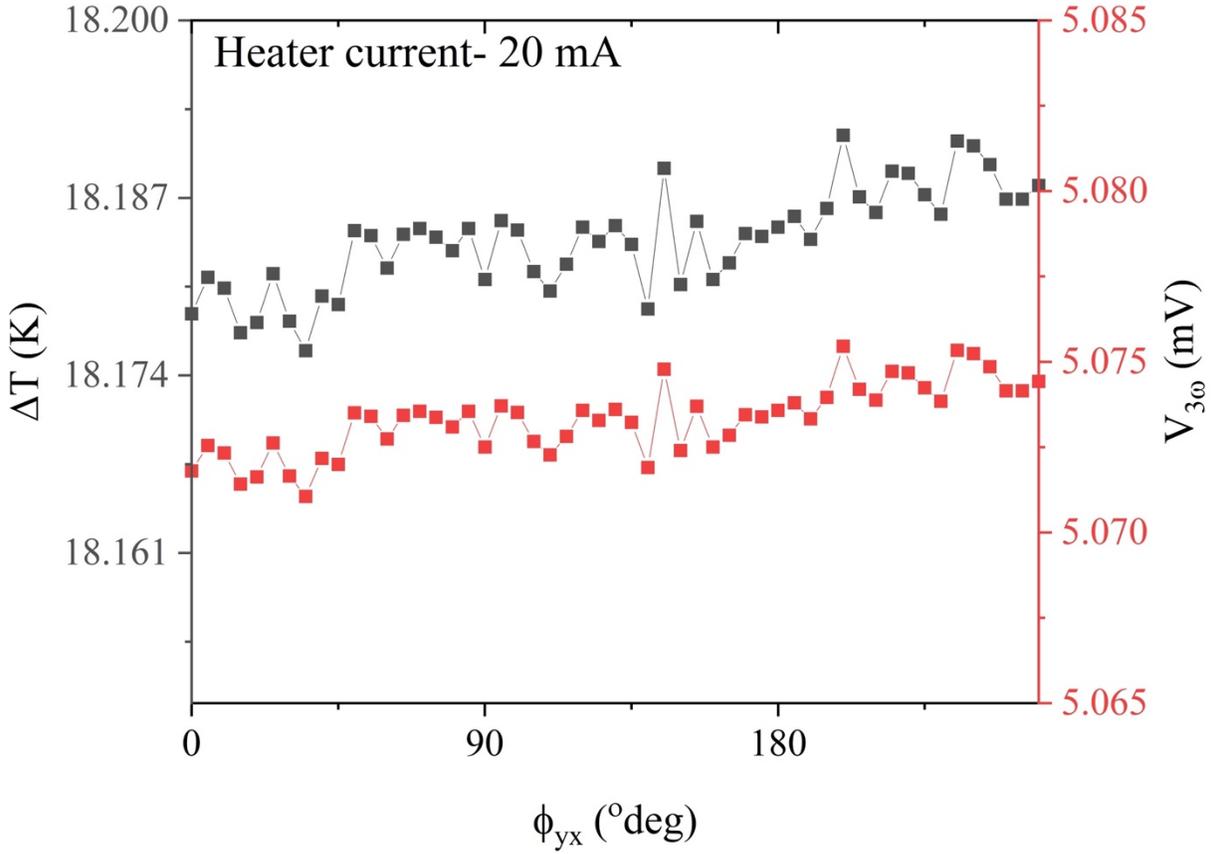

Supplementary Figure S5. The estimated increase in heater temperature and the $V_{3\omega}$ response at an applied magnetic field of 2 T.

### D. The effect of heating current on transverse thermoelectric response in 50 nm and 5 nm a-Si samples

We measured the $V_{2\omega}$ response as a function of magnetic field in IM configuration in $Ni_{80}Fe_{20}$/a-Si (50 nm) sample at 20 mA of heating current. The total response was 33.51 µV. Similarly, we also measured the angle dependent $V_{2\omega}$ response $Ni_{80}Fe_{20}$/a-Si (5 nm) sample. The angle dependent response demonstrated the symmetry expected for SSE/ANE behavior. In addition, the overall response was measured to be 67.5 µV at 20 mA of heating current.

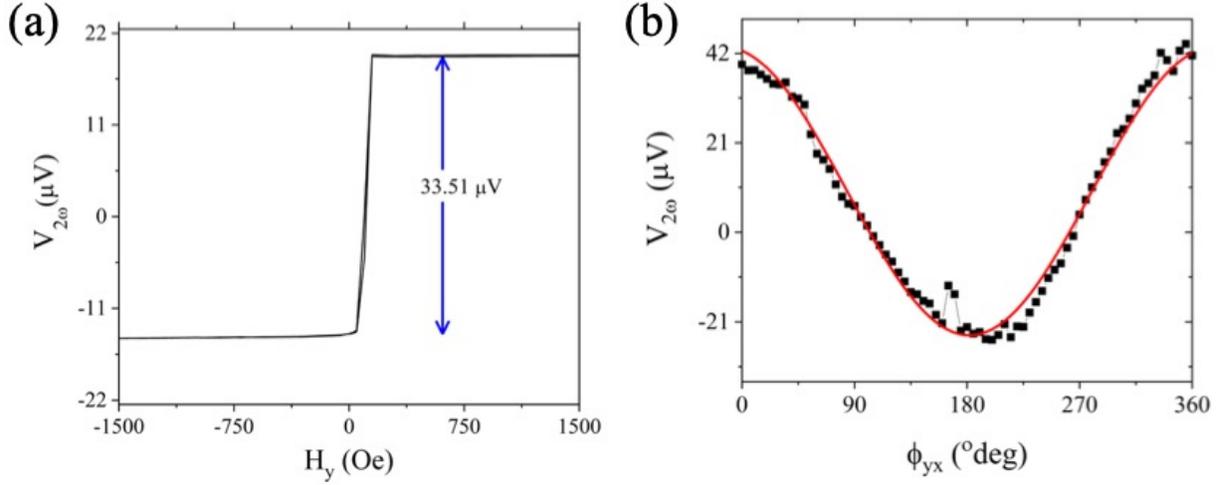

Supplementary Figure S6. (a) The $V_{2\omega}$ response as a function of magnetic field in IM configuration in $Ni_{80}Fe_{20}$/a-Si (50 nm) sample and (b) the angle dependent $V_{2\omega}$ response $Ni_{80}Fe_{20}$/a-Si (5 nm) sample showing the expected cosine symmetry in yx-plane.

### E. Magneto-thermal measurement in unstrained samples with heater at the bottom

We had hypothesized that strain gradient mediated Rashba SOC is the underlying cause of large spin dependent thermal responses presented in this study. Thus, these responses should disappear if the strain is removed. We demonstrated this situation by modifying the experimental setup – switching the position of the heater and the sample, as shown in Supplementary Figure S6. In this configuration, the sample is no longer constrained by the MgO (insulator) and Pt (heater) layers, thus, the strain and strain gradient effects will be significantly reduced. We fabricated set of devices in the new configuration having the following structure- a-Si (50 nm)/$Ni_{80}Fe_{20}$ (25 nm), $Ni_{80}Fe_{20}$ (25 nm) and Pt (3 nm)/$Ni_{80}Fe_{20}$ (25 nm). We will call the new samples as unstrained while the samples in the main text will be referred to strained.

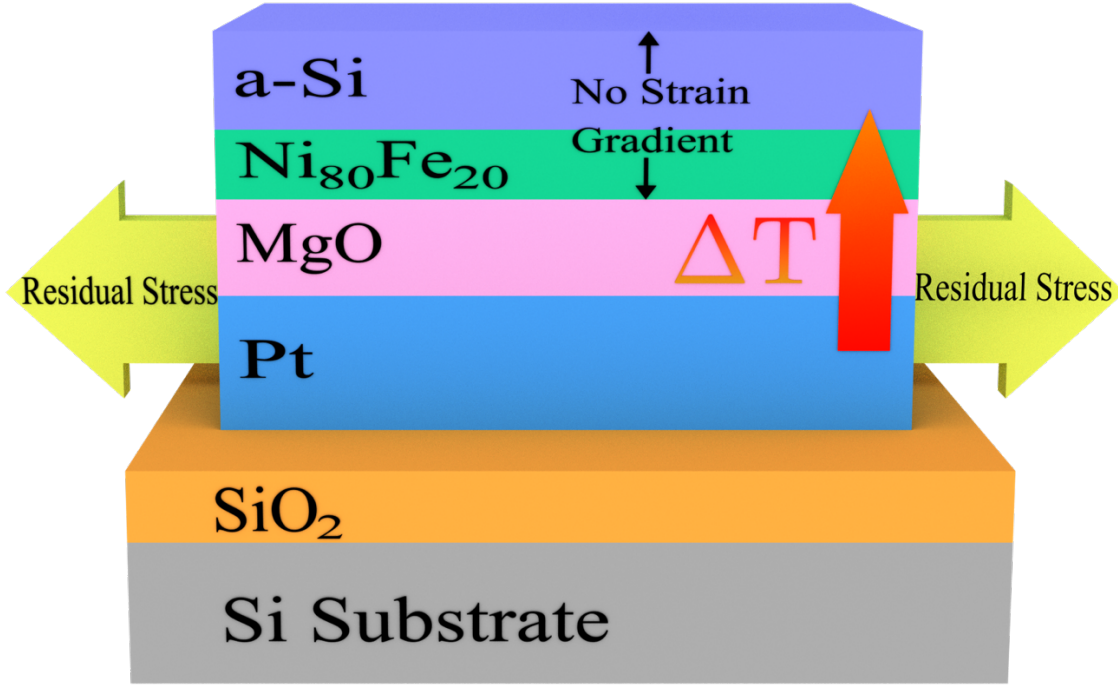

Supplementary Figure S7. The schematic showing the experimental setup for unstrained samples.

The transverse thermoelectric response in IM configuration was 0.95 µV, 0.35 and 0.71 in a-Si (50 nm)/$Ni_{80}Fe_{20}$, $Ni_{80}Fe_{20}$ and Pt/$Ni_{80}Fe_{20}$ samples, respectively, as shown in Supplementary Figure S7 (a)-(d). The transverse thermoelectric response in case of Pt/$Ni_{80}Fe_{20}$ was estimated by angle dependent measurement as shown in Supplementary Figure S8 (d) since planar Nernst effect (PNE) response made it difficult to measure in field dependent measurement as shown in Supplementary Figure S7 (c). Similar to strained samples, the SSE response in Pt (0.36 µV) was estimated to be same as the ANE (easy axis) response in $Ni_{80}Fe_{20}$ (0.35 µV). The spin dependent thermal response in a-Si sample was 1.67 times larger than that of Pt sample as compared to four times in strained sample, which confirmed the effect of strain on SSE. Based on the ANE (easy axis) coefficient of 0.1 µV/K reported earlier, we estimated the temperature difference of 0.27 mK across the thickness of the sample. Hence, ANE (easy axis) coefficient, estimated in this study, is believed to be correct for our experimental configuration.

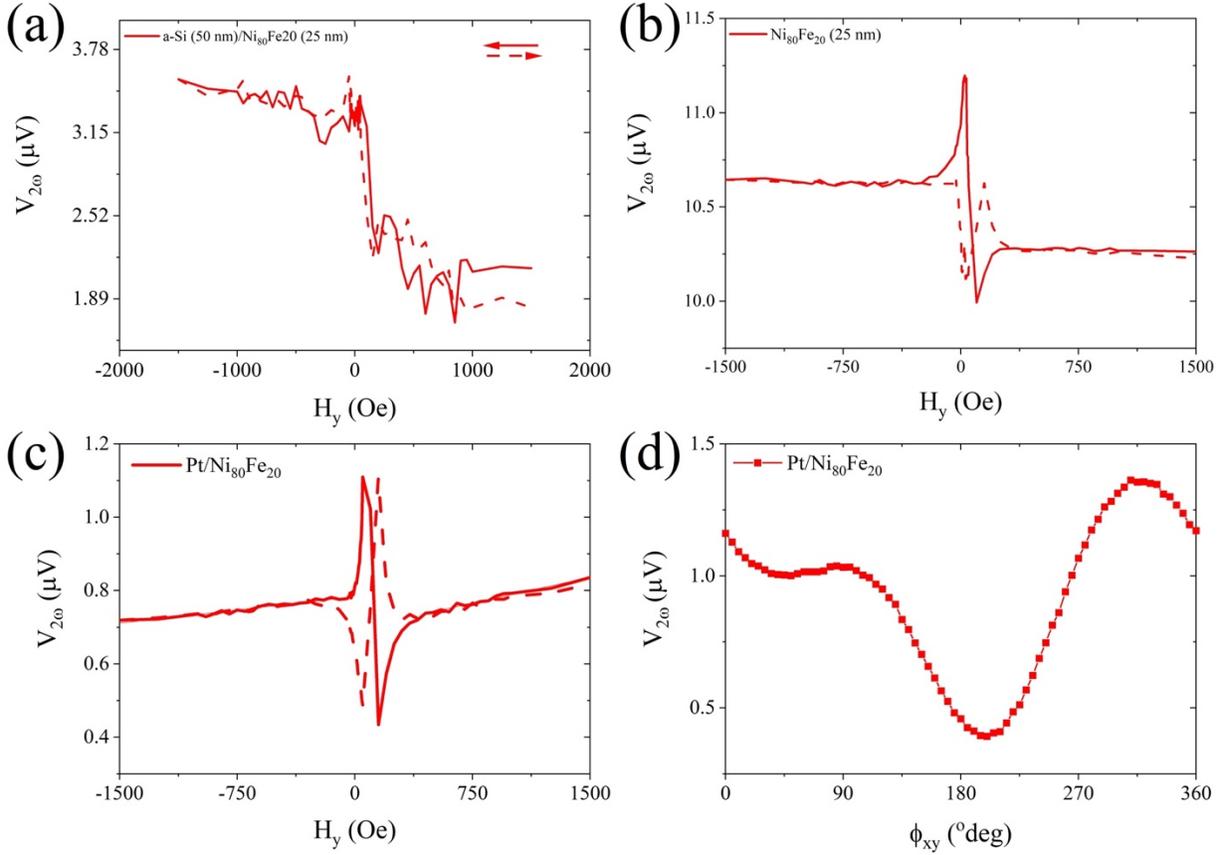

Supplementary Figure S8. The magnetic field dependent transverse thermoelectric response measurement in PM configuration in (a) a-Si (50 nm)/ $Ni_{80}Fe_{20}$ (25 nm), (b) $Ni_{80}Fe_{20}$ (25 nm) and (c) Pt (3 nm)/$Ni_{80}Fe_{20}$ (25 nm) samples, respectively. (d) the angle dependent (in xy-plane) transverse thermoelectric response in Pt (3 nm)/$Ni_{80}Fe_{20}$ (25 nm) sample.

We, then, measured the transverse thermoelectric response in PM configuration. The measured responses are shown in Supplementary Figure S8. The $Ni_{80}Fe_{20}$ sample showed a hard-axis ANE behavior and measured response is 15.45 µV. The Pt sample also exhibits hard-axis ANE behavior and measured response is 15.4 µV. Using the planar Nernst response ($\sin 2\theta$ in xy-rotation as shown in Supplementary Figure S8 (d) of 0.21 µV and PNE coefficient of 70 nV/K, we estimated a $\Delta T_y$= 0.1875 K along the width of the sample. The corresponding out of plane $2V_{ANE}$ was 15.4 µV. Using the temperature information, we estimated the ANE coefficient for in-plane temperature gradient:

$$S_{ANE} = \frac{V_{ANE} w}{L \Delta T} \qquad (S3)$$

where, $2V_{ANE} = V_{2\omega} = 15.4$ µV, w= 10 µm, L = 160 µm and $\Delta T$ is the temperature difference across the width of $Ni_{80}Fe_{20}$ thin film. The ANE coefficient for hard-axis magnetization was estimated to be 2.565×10$^{-6}$ μV/K. Using the slope of high field behavior, we could also estimate the ordinary Nernst effect (ONE) coefficient. The unstrained samples had the slopes of 0.338 μV/T, 0.278 μV/T and 0.0015 μV/T in Ni$_{80}$Fe$_{20}$, Pt/ Ni$_{80}$Fe$_{20}$ and a-Si (50 nm)/ Ni$_{80}$Fe$_{20}$ samples, respectively. Using $\Delta T_y$ = 0.1875 K, we estimated that the S$_{ONE}$ were 0.113 μV/(KT) and 0.0926 μV/(KT) in Ni$_{80}$Fe$_{20}$ and Pt/ Ni$_{80}$Fe$_{20}$ samples, respectively, as shown in Table 2 (main text). For the unstrained a-Si (50 nm)/ Ni$_{80}$Fe$_{20}$ sample, we assumed that the ANE coefficient would be same as Ni$_{80}$Fe$_{20}$. The resulting S$_{ONE}$ in unstrained a-Si sample was estimated to be 0.0015 μV/(KT).

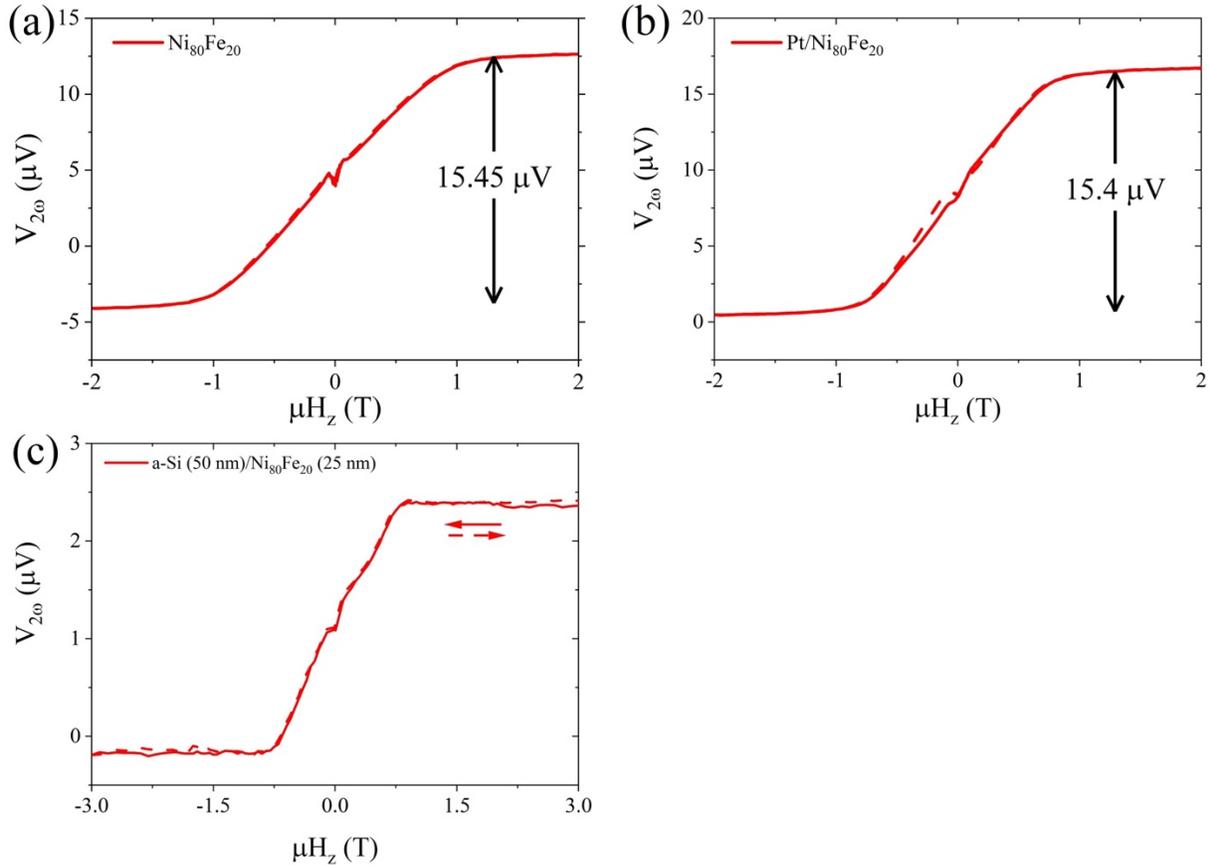

Supplementary Figure S9. The magnetic field dependent transverse thermoelectric response measurement in PM configuration in (a) Ni$_{80}$Fe$_{20}$ (25 nm), (b) Pt (3 nm)/Ni$_{80}$Fe$_{20}$ (25 nm) and (c) a-Si (50 nm)/ Ni$_{80}$Fe$_{20}$ (25 nm) samples, respectively.

We, then, measured the slope for ONE in strained samples– +0.445 μV/T, +2.13 μV/T, -8.3 μV/T/6.3 μV/T and -3.6 μV/T in Ni$_{80}$Fe$_{20}$, Pt/ Ni$_{80}$Fe$_{20}$, Ni$_{80}$Fe$_{20}$/a-Si (5 nm) and Ni$_{80}$Fe$_{20}$/Cu/a-Si (25 nm) as shown in Figure 3 of main text. In Ni$_{80}$Fe$_{20}$ sample (strained), we estimated the horizontal temperature difference using ANE and ONE coefficient's to be 0.64 K and 0.247 K, respectively, as shown in Table 2. However, the ONE coefficient reflects primarily the bulk behavior whereas ANE response can be significantly affected by the interfaces. Hence, the temperature difference calculated using ONE behavior is assumed to be closer to actual temperature difference. Using $\Delta T_y$=0.247 K and ANE (hard axis) response of 52.5 µV, we estimated the $S_{ANE}$ in strained $Ni_{80}Fe_{20}$ sample to be 6.6×10$^{-6}$ µV/K instead of 2.565×10$^{-6}$ µV/K in the unstrained sample. It is noted that unstrained sample had one interface with MgO while the strained sample had two interfaces one with MgO and other with $SiO_2$. And this increase in ANE (hard axis) coefficient was expected to arise due to spin dependent behavior at interfaces with oxides. To demonstrate the effect of interface, we compared the ONE coefficient's in strained and unstrained Pt devices and estimated the temperature a difference of $\Delta T_y$=1.437 K in strained Pt device. Using this temperature information, we estimated the ANE (hard axis) coefficient from 185 µV ANE (hard axis) response to be 4.65×10$^{-6}$ µV/K, which was smaller than strained $Ni_{80}Fe_{20}$ sample. This difference in ANE (hard axis) coefficients was attributed to the absence of MgO/$Ni_{80}Fe_{20}$ interface in the strained Pt/$Ni_{80}Fe_{20}$ sample. For 5 nm a-Si strained sample, the ONE response was negative and could not be considered to be same as $Ni_{80}Fe_{20}$ sample in spite of a-Si being only 5 nm. Instead, we estimated the temperature difference of $\Delta T_y$=0.726 K from 95.93 µV of ANE (hard axis) response and using ANE (hard axis) coefficient of 6.6×10$^{-6}$ µV/K. Then, the corresponding $S_{ONE}$ would be -0.542 µV/(KT) and -0.714 µV/(KT) for negative and positive magnetic fields, respectively. The a-Si sample clearly showed an asymmetry in the ONE measurement. Using similar assumption, we estimated the $\Delta T_y$=0.338 K and $S_{ONE}$=-0.664 µV/(KT) in $Ni_{80}Fe_{20}$/Cu/a-Si (25 nm) sample. This exercise demonstrated approximate temperature differences and corresponding ANE (hard-axis) and ONE coefficients, which are consistent with values reported in literature.

Additionally, in case of unstrained a-Si (50 nm)/$Ni_{80}Fe_{20}$ sample, the transverse thermoelectric response in PM configuration exhibits a hard-axis ANE behavior and response is ~2.5 µV as shown in Supplementary Figure S9 (c). This behavior is opposite as compared to the strained sample, as shown in Figure 3 (c) in the main text. In strained sample, the magnitude of low field response in PM configuration is similar to total response in IM configuration, which means that 0.95 µV of low field thermal response should have been present in PM configuration for unstrained sample. The absence of such response clearly indicated that the observed behavior arose due to strain and strain gradient at the interface. In addition, the response in PM configuration also proves that the interstitial Cu and Ni atoms are not the primary drivers of spin dependent behavior reported in this study. Instead, interfacial roughness in conjunction with strong Rashba SOC might be the underlying cause of observed behavior PM configuration as hypothesized.